\documentclass[conference]{IEEEtran}
\IEEEoverridecommandlockouts
\usepackage{cite}
\usepackage{amsmath,amssymb,amsfonts}
\usepackage{algorithmic}
\usepackage{graphicx}
\usepackage{textcomp}
\usepackage{xcolor}
\usepackage{soul}
\def\BibTeX{{\rm B\kern-.05em{\sc i\kern-.025em b}\kern-.08em
    T\kern-.1667em\lower.7ex\hbox{E}\kern-.125emX}}
\begin{document}

\title{On Coding for Reliable VNF Chaining in DCNs \\
\thanks{This work has been performed in the framework of
SENDATE-PLANETS (Project ID C2015/3-1), and it is partly
funded by the German BMBF (Project ID 16KIS0470).
}
}

\author{\IEEEauthorblockN{1\textsuperscript{st} Anna Engelmann}
\IEEEauthorblockA{\textit{TU Braunschweig}\\
Braunschweig, Germany \\
a.engelmann@tu-bs.de}
\and
\IEEEauthorblockN{2\textsuperscript{nd} Admela Jukan}
\IEEEauthorblockA{\textit{TU Braunschweig}\\
Braunschweig, Germany \\
a.jukan@tu-bs.de}
\and
\IEEEauthorblockN{3\textsuperscript{nd} Rastin Pries}
\IEEEauthorblockA{\textit{Nokia Bell Labs}\\
Munich, Germany \\
rastin.pries@nokia-bell-labs.com}
}
\maketitle

\begin{abstract}
We study how erasure coding can improve service reliability in Data Center Networks (DCN). To this end, we find that coding can be best deployed in systems, where i) traffic is split into multiple parallel sub-flows, ii) each sub-flow is encoded; iii) SFC along with their corresponding Virtual Network Functions (VNF) concatenated are replicated into at least as many VNF instances as there are sub-flows, resulting in parallel sub-SFCs; and iv) all coded sub-flows are distributed over parallel paths and processed in parallel. We study service reliability as function of the level of parallelization within DCN and the resulting amount of redundancy. Based on the probability theory and by considering failures of path segments, VNF and server failures, we analytically derive the probability that parallel sub-flows are successfully processed by the parallelized SFC and that the original serial traffic can be successfully recovered without service interruptions. We compare the proposed failure protection with coding and the standard backup protection and evaluate the related overhead of both methods, including decoding, traffic redirection and VNF migration. The results not only show the benefit of our scheme for reliability, but also a reduced overhead required in comparison to backup protection. 
\end{abstract}

\begin{IEEEkeywords}
VNF, SFC, Reliability, Coding, Combinatorial analysis, Parallelism
\end{IEEEkeywords}

\section{Introduction}
Network Function Virtualization (NFV) enables a virtualization of traditional network functions by replacing them with Virtual Network Functions (VNFs) allocated typically in virtual machines (VMs) or containers in data center networks (DCN). In NFV-based DCNs, the failure of any VNF within the service function chain (SFC) disrupts the entire chain causing service interruptions. An VNF failure is typically a result of hardware (server) or software (VM) failures, or connectivity loss (failure of path segments). To achieve the required level of SFC reliability, the backup-based VNF protection is utilized. To be effectively used, a backup protection requires fast failure detection and fast reaction of detected failures, to minimize the traffic losses. At the same time, it requires VNF migration and traffic redirection\cite{REL-ETSI, Herker:2015,Qu:2016,Ding:2017, Hmaity:2016, Ye:2016}.

\par In this paper, we study how instead of backup protection, a \emph{systematic erasure coding} can improve service reliability in DCNs, while alleviating or even eliminating the challenges associated with VNF failure detection, fast VNF migration and traffic redirection. To this end, we find that coding can be best deployed in systems, where  traffic is split into multiple parallel sub-flows (parallelized). Since DCNs usually deploy Equal Cost Multipath Protocol (ECMP), which breaks large flows into parallel sub-flows and uniformly distributes the resulting sub-flows over the network \cite{Xu:2014, Chakraborty:2016, Engelmann:ICC18}, the idea of deploying traffic parallelism for reliability is also practically valid. The novelty of our system is that we encode each sub-flow and replicate VNFs of each SFC along, resulting in parallel sub-SFCs. Finally, all coded sub-flows are distributed over multiple parallel paths and processed in parallel.  

This paper presents an analytical study of the impact of coding on SFC reliability in traffic-, flow- and path-parallelized DCNs. In case of VNF failures, we show that there is no need for complex failure detection, time consuming VNF migrations and traffic redirection. We compare VNF failure protection with systematic erasure coding and backup failure protection. Based on the probability theory and the reliability of the individual VMs, servers and path segments, we derive analytical expressions for evaluating the SFC (service) reliability. We define the overhead of backup protection as the need for VNF migration and traffic redirection, while the overhead of coding protection as a need for the computationally intensive decoding. The results show that, compared to backup protection, the coding protection can provide higher service reliability and lower probability for decoding than probability for VNF migration and traffic redirection. To the best of our knowledge, this is the first study to analyze a suitability of erasure coding to improve the service reliability in DCNs.

\par The rest of the paper is organized as follows. Section 2 presents the reference network architecture. The methods for VNF failure protection are drawn in Section 3. Section 4 presents the reliability analysis. Numerical results are shown in Section 5. Section 6 concludes the paper.
 \begin{figure}[!t]
\centering
\includegraphics[width=1\columnwidth]{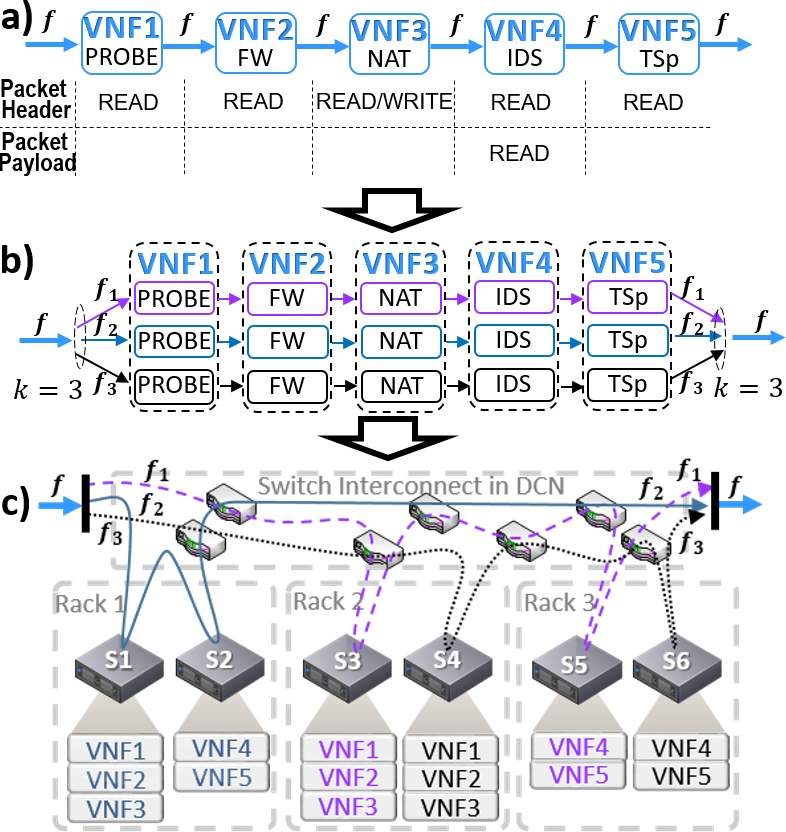}%
  \vspace{-0.1cm}
  \caption{a) Serial DCN traffic: traffic flow $f$ is sent over one path through one SFC; b) Parallel VNFs with replications and parallel SFCs: Traffic flow $f$ is parallelized into $k=3$ sub-flows ($f_1$, $f_2$, $f_3$); SFC is parallelized into $k=3$ sub-SFCs; parallel sub-flows are sent over $k$ parallel paths through $k$ parallel sub-SFCs. c) Parallelism in DCN and VNF placement: parallel sub-SFCs are distributed over $N=2$ servers each and disjoint from each other; parallel sub-flows are sent over parallel disjoint paths and different servers.}
\vspace{-0.6cm}
  \label{par}
\end{figure}
\section{System Model}

\subsection{DCN model}
The DCN hardware fabric consists of racks with multiple servers, links and two types of switches: Top-of-Rack (ToR) and intermediate forwarding switches. ToR switches provide connectivity to every server within the rack and to every intermediate switch in the DCN. The intermediate switches are connected to all ToRs and perform forwarding across racks. We assume that the traffic towards other racks is routed by the source ToR, while the intermediate switches just forward the traffic between racks based on concepts of segment routing (SR) \cite{CORD}. For example, the source ToR in the source rack addresses the destination ToR in the destination rack by packet labeling and sends traffic to a certain intermediate switch. The intermediate switches perform only label lookups to route the traffic to the destination ToR. SR allows to change an end-to-end path utilized over the network, whereby the source ToR switch needs only to change the label. This is an important capability of a DCN network in case of backup VNF protection.

We assume that each server can host multiple VMs and a virtual switch (vS), e.g., programmable hypervisor switch. Each VM allocates one VNF only. All VMs are connected to a vS, which selects one VNF for processing of arrived packets and provides a connectivity to the ToR switch. 
The ordered sequence of VNFs builds an SFC. Fig. \ref{par}a) shows an example of an SFC, which consists of 5 different VNFs allocated in DCN. Different network functions read/write different parts of received packet during its operation, header or payload or both. We refer to VNFs, which read or write packet header only, as \emph{header-VNFs} (h-VNF) and to VNFs, which read or write packet payload, as \emph{payload-VNFs} (p-VNF). The most of VNFs considered in the example are h-VNFs, while only NAT changes the header and only IDS is p-VNF, i.e., looks at payload. To complete the service, all packets need to be processed in order. In our example, all incoming packets are first registered by passive probing function (Probe), then Firewall (FW) proves if traffic is allowed to pass through DCN. The legitimate packets from FW are processed by Network Address Translation (NAT), which replaces the IP addresses and port numbers in packet header. After that, packets are sent to an Intrusion Detection System (IDS), which copies the flow to perform offline traffic analyses to identify and log violations. Finally, in the Traffic Shaper (TSp), the packets are categorized and queued to meet a certain quality of service (QoS). 

 \begin{figure*}[!t]
\centering
\includegraphics[width=1.46\columnwidth]{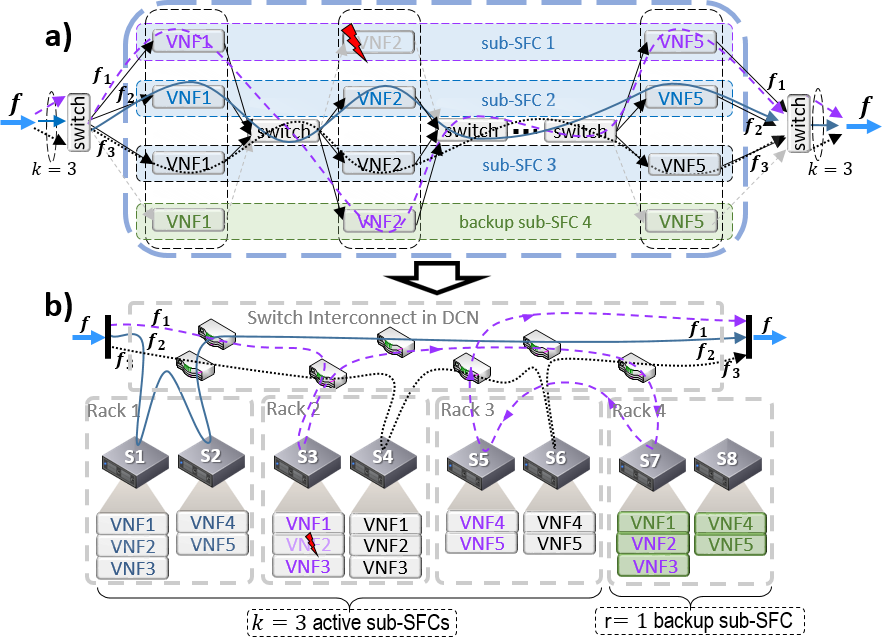}%
  \vspace{-0.1cm}
  \caption{Backup protection: a) Network view: SFC is parallelized into $k=3$ active and $r$ redundant backup sub-chains; switches inside DCN can redirect any sub-flow from failed VNF to backup VNF of the same type; b) Server view: active and backup sub-SFCs are distributed over $N=2$ servers, sub-flows $f_1$ is redirected to backup servers S7 and S8, respectively; backup VNF3 is activated to reduce latency.}
\vspace{-0.6cm}
  \label{backupPlace}
\end{figure*}

\subsection{Parallelism in DCN}
Without loss of generality, let us simplify the definition of \emph{traffic flow} or \emph{flow} as a large number of IP packets with the same IP addresses and port numbers in the header. In this section, we discuss in detail three following concepts that we apply to DCN parallelism:  1) Traffic Parallelism: a serial traffic flow is split into a number of parallel sub-flows at the sender; 2) SFC Parallelism: all VNFs of a certain SFC are \emph{replicated} into as many VNF instances as there are sub-flows, resulting in parallel sub-SFCs; and 3) Path Parallelism: all parallel sub-flows are independently transmitted over link disjoint paths and processed in parallel by parallel disjoint sub-SFCs, distributed over different servers. 

\subsubsection{Traffic Parallelism}
Traffic spreading is a well known technique to improve load balancing on the network interfaces \cite{Xu:2014, Chakraborty:2016}. As shown in Fig. \ref{par}b), any traffic flow $f$ can be split (parallelized) into $k=3$ sub-flows, i.e., $f_1$, $f_2$ and $f_3$. To realize traffic parallelism via traffic spreading, however, the incoming traffic needs to be sorted into different flows based on source and destination IP addresses and port numbers in packet headers. This is because different traffic flows would require different SFCs. Finally, packets of any certain flow $f$ can be distributed over $k$ interfaces in Round Robin fashion building $k$ parallel sub-flows. When outgoing DCN traffic needs to be sent in serial fashion, after service completion by the certain SFCs, all packets from all flows and sub-flows need to be serialized into one data stream based, for instance, on FIFO. 

\subsubsection{SFC Parallelism}
Generally, parallel sub-flows can be either sent to the same VNF instances of an SFC, or a separate replicas of an SFC, consisting of certain VNFs, needs to be created for each sub-flow. For the latter, which we call SFC parallelism, all VNFs of certain SFC are replicated into $k$ VNF instances resulting in $k$ parallel sub-SFCs. As a result, each sub-flow passes its own sub-SFC as shown in Fig. \ref{par}b). This can be implemented with SFC Encapsulation as proposed in \cite{RFC7665}, whereby each parallel sub-SFC can be characterized as Service Function Path with special ID. For that reason, each packet requires an additional Network Service Header with Service Path Identifier of 24 bits as discussed in \cite{RFC8300}. 
It is necessary that all $k$ VNF replicas of certain type process packets of certain flow based on same rules and with same processing result. For instance, all $k$ NAT replicas in Fig. \ref{par}b) have to replace IP addresses and ports in the packet headers independent of sub-flow by the same predefined IP addresses and ports. Thus, the parallel VNFs need some kind of management unit for its coordination and rule update. When flow and SFC parallelism are deployed, the synchronization between VNFs of the same type can be provided by an external state repository, which can store internal states of VNFs \cite{REL-ETSI}.

\subsubsection{Path Parallelism}
In general, to reach better reliability, VNFs from different sub-SFCs should not be placed in the same server. This prevents the usage of the same links and servers by parallel sub-flows and, thus, additionally increases reliability. Thus, after traffic parallelization, the $k$ parallel sub-flows should be sent over $k$ parallel link and server disjoint paths, i.e., through $k$ disjointedly placed sub-SFCs. Then, each parallel sub-flow passes through its own sub-SFC, which is distributed over $N$ servers. Fig. \ref{par}c) illustrates a possible sub-SFCs placement within DCN, whereby VNFs of any sub-SFC for a certain sub-flow are distributed over $N=2$ servers and disjoint from other sub-SFCs, e.g., sub-SFC of black sub-flow $f_3$ is placed in servers S4 and S6 and disjoint from other sub-SFCs (blue and purple) placed in S1-S3 and S5. Thus, the VNFs of the same type are placed in different servers, e.g., VNFs1, VNFs2 and VNFs3, are distributed over servers S1, S3 and S4.  Any flow is successfully processed by certain SFC, when all parallel sub-flows passed its sub-SFC and meet in the same server to be serialized to leave DCN. 

\begin{figure*}[!t]
\centering
\includegraphics[width=2\columnwidth]{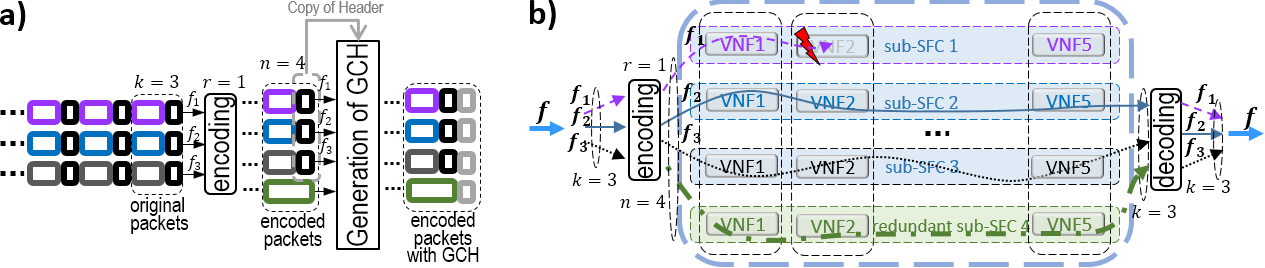}%
  \vspace{-0.1cm}
  \caption{Protection with systematic erasure coding: a) packet encoding process and generation of GCH; b) VNF deployment in case of VNF failures. The $3$ parallel sub-flows are encoded into $4$ parallel sub-flows and SFC is parallelized into $4$ sub-SFCs. Encoded sub-flows are sent over disjoint paths to disjoint servers through disjoint sub-SFCs. The loss of main sub-flow $f_1$ due to failure of VNF2 of sub-SFC1 results in decoding need.}
\vspace{-0.6cm}
  \label{codingUse}
\end{figure*}
\subsection{Failure model}
We consider three type of failures, i.e., failures of servers, VNFs and connections between servers (path segments). The switches in DCN are assumed as highly reliable and never fail. Since we assume that each VM reserves as many resources as required for one VNF to serve incoming packets, the failure of any VM causes failure of one only VNF. The server failure, in contrast, causes failure of all VMs running on this server and, thus, failure of multiple VNFs. Finally, any failure of path segments results in no connectivity to a certain server.  This is equivalent to the server failure and failure of all VNFs running on that server. As shown in Fig. \ref{par}c), each sub-SFC can be allocated in two different servers, each path for each sub-flow consists of three path segments, i.e., 9 disjoint path segments totally, any of which can fail. A failure of path segments between S1 and S2 is a failure of server S2 and VNF4 and VNF5 of second (blue) sub-SFC. Usually any VNF failure, e.g., due to VM, server or path segment failures, requires a new VNF instantiation or VNF migration and traffic redirection. 

\section{Failure Protection}
A parallelized SFC is characterized by a certain level of reliability, which depends on the reliability of the underlying VNFs, servers and  path segments. To increase the SFC reliability, we next consider two protection methods: A)  backup resources; and B) systematic erasure coding.

\subsection{Backup protection}
To protect  active h-VNFs and p-VNFs over all parallel sub-SFCs, a parallelized SFC can use a parallel backup sub-SFCs. In case of VNF failure, the backup VNFs of backup sub-SFC can replace any failed VNF of the same type over any sub-flow. According to \cite{REL-ETSI, Herker:2015}, to provide reliable communications, the most effective VNF deployment is to utilize multiple switches inside the DCN between individual VNFs, such as vS, ToR and intermediate switches, which can detect failure and redirect traffic to the corresponding available backup VNF, when the connection to an active server or VNF fails. Fig. \ref{backupPlace}a) illustrates the idea, whereby SFC is parallelized into $k=3$ active and $r=1$ backup sub-SFCs. The $k=3$ parallel sub-flows are sent over $3$ parallel disjoint paths through $3$ sub-SFCs. Since VNF2 of sub-SFC1 fails, the switch inside the DCN redirects sub-flow $f_1$ to VNF2 of backup sub-SFC $4$. 

Fig. \ref{backupPlace}b) shows a possible placement and deployment of backup sub-SFC in DCN. Just like an active sub-SFCs, a backup sub-SFC is distributed over $N=2$ servers in Rack4 providing the most reliable placement of backup VNFs. In this example, active VNF2 of sub-flow $f_1$ fails. The sub-flow $f_1$ is redirected to backup server S7, where VNF2 and VNF3 are activated to replace original VNF2 and VNF3 from server S3. The backup VNF3 on server S7 is activated to avoid sending $f_1$ back to S3 and, thus, additional delays. As a result, $f_1$ is processed by backup VNF2 and VNF3 on S7 and, then, sent to S5 to complete the service by VNF4 and VNF5 in Rack 3.
 \begin{figure*}[!th]
\centering
\includegraphics[width=2\columnwidth]{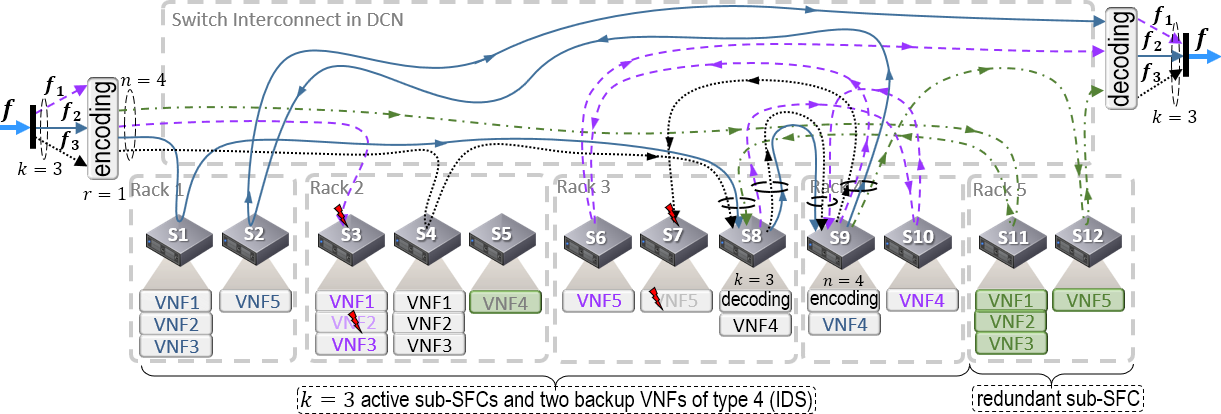}%
  \vspace{-0.2cm}
  \caption{Deployment and placement of sub-SFCs in DCN in case of Hybrid protection: Coded sub-flows need to be decoded prior to processing by VNF4.}
 \vspace{-0.5cm}
  \label{hybridPlace}
\end{figure*}
\subsection{Protection with erasure coding}
The use of Forward Error Correction (FEC) codes such as \emph{erasure coding} is a classical solution to improve the reliability of multicast and broadcast transmissions.
Erasure codes can add redundancy into data to protect it against losses.  Any ($n$, $k$) erasure code encodes $k$ units of original data into $n$ units of coded data, in which any $k$ out of $n$ units can recover the original data and, thus, the code can tolerate the failures of up to $r=n-k$ data units.  We refer to any $k$ units of original data encoded together as \emph{generation} and to $r$ additional data units generated by encoding as \emph{redundancy}.
Most erasure codes deployed in practice are \emph{systematic codes}, meaning that the $k$ original data units are unchanged after encoding and only the $r$ redundant data units present a linear combination of $k$ original data units. Thus, that $k$ data units of coded data can be directly accessed without prior decoding. In contrast, when at least one out of $k$ unchanged data units is lost, any out of $r$ redundant units from the same generation replaces the lost data unit, while decoding is required to recover $k$ original data units. The decoding is only successful, if at least $k$ data units from each generation arrive at the decoder. For VNF protection with coding, we assume that each parallel sub-flow can be presented as a sequence of data units, whereby the original and encoded data units have the same predefined size. We refer to the $k$ parallel sub-flows, which are unchanged after encoding, as \emph{main sub-flows}, and to the sub-flows, which are built by redundancy as \emph{redundant sub-flows}.

To be able to recover both header and payload, we propose to encode both, i.e., the whole packets, which are interpreted as a data unit by encoder. An example of encoding process is illustrated in Fig. \ref{codingUse}a, where $k=3$ parallel sub-flows encoded to generate $r=1$ redundant sub-flow. The three original packets, one from each parallel sub-flow, are built generation and combined together into the green redundant packet. In contrast to original packets, which are unchanged after encoding, the redundant packet does not have any header or readable payload, while belonging now to the same flow as original packets and needs to pass through same chain of VNFs. Usually, VNFs are only able to process uncoded data. Thus, there is a need for new header for generated redundant packets in accordance with original header. Thus, we propose to utilize a \emph{generalized coding header} (GCH).  As presented in Fig. \ref{codingUse}a, the original headers are copied to generate a certain GCH with the same source/destination addresses and the same port numbers as determined for the original traffic flow. The generator of GCH (Generation of GCH) utilize a copy of original IP and TCP headers and change some header fields: 1) In the IP-GCH, the Header Length, Total Length and Header Checksum are adapted to the coded packet length; 2) In the TCP-GCH, data offset and header checksum needs to be changed. Moreover, the sequence numbers need to be redefined to identify the packets belonging to the same generation.  Based on this new sequence number in TCP-GCH, the decoder can recognize packets from the same generation required for successful flow recovery or decoding. Finally, the GCHs (gray) are attached to the certain encoded packet. Additionally, the metadata can be inserted to each packet to hold information related to coding process, e.g., the number of main and redundant sub-flows, generation size, etc.  
As a result, any h-VNF can process (read/write) GCH and, thus, coded redundant packets. The information from GCH is copied to the original header after any flow recovery or any decoding. Without loss of generality, the encoding (decoding) process requires packet buffering and clocking to build generations. That can be implemented by VNFs such as Traffic Shaper.

Fig. \ref{codingUse}b shows the VNF protection by systematic erasure code, where the original flow $f$ is parallelized into $k=3$ sub-flows. The parallel sub-flows are encoded into $n=4$ sub-flows providing $r=1$ redundancy. That requires $r=1$ redundant sub-SFCs.   
In the example, there are $k=3$ main sub-flows after encoding, $f_1$, $f_2$ and $f_3$, while redundant sub-flow presents a linear combination of them. After encoding, the sub-flows are sent over $n=4$ disjoint paths through $n=4$ parallel sub-SFCs. The VNF2 of sub-SFC1 fails, resulting in loss of the main sub-flow $f_1$. The blue ($f_2$), black ($f_3$) and green sub-flows ($k=3$) arrive at destination, whereby the recovery of original flow is possible by decoding. 
The decoding is only required, when at least one out of $k$ main sub-flows is lost. However, the loss of at most $r$ sub-flows can be tolerated. In contrast to the presented example in Fig. \ref{codingUse}b, the decoding is not required, if redundant (green) sub-flows were lost.



Any SFC generally consists of both  h-VNFs and p-VNFs, which process packet header and packet payload, respectively. In contrast to h-VNFs, which are processing packet header and, thus, can work with GCH, p-VNFs such as IDS can not work with coded packets and require recovery of original sub-flows or decoding prior to processing. As a results, p-VNFs can be protected by backup only. Then, we propose to use a hybrid backup-coding protection of SFC, which consists of both h-VNFs and p-VNFs. An example of hybrid VNF protection is presented in Fig. \ref{hybridPlace}, where VNF4 is only a p-VNF, which can need prior decoding and backup protection. 
After traffic parallelization into $k=3$ original sub-flows, i.e., $f_1$, $f_2$ and $f_3$ and encoding them into $n=4$ coded sub-flows.  All $n=4$ sub-flows are sent over disjoint paths through different sub-SFCs. The purple sub-flow ($f_1$) is lost due to failure of connection to S3 or failure of S3 or VNF2 failure. The $3$ remaining sub-flows (blue, black and green) are successfully processed by the first three h-VNFs of related sub-SFCs in servers S1, S4 and S11. Since the main purple sub-flow is lost, the decoding is required in S8 to recover all $3$ original sub-flows (purple, blue and black) prior to processing by p-VNF4 (IDS). During decoding, the green redundant sub-flow replaces the lost purple sub-flow. The recovered original sub-flows are sent through different instances of VNF4 allocated in servers S8, S9, S10. The $r=1$ backup VNF4 in server S5 can replace any failed active VNF4 requiring sub-flow redirection. Finally, the original sub-flows are encoded in S9 into $n=4$ sub-flows and sent to h-VNFs of type 5. The black VNF5 in server S7 fails resulting in the lost of the main black  sub-flow and a need for decoding at destination prior to traffic serialization.
\section{Analysis}

The analytical models for availability and reliability for complex systems are interchangeable and only defined through availability, or reliability, of the individual components \cite{REL-ETSI}. Thus, we refer to the end-to-end service availability (reliability) as \emph{service success}, defined as a probability that at least $k$ parallel sub-flows of certain traffic flow can successfully traverse $k$ parallel sub-SFCs and be serialized at end-hosts. We summarized the notations used in Table \ref{t1}. We assume that each sub-SFC consists of $\Psi$ different VNFs and each sub-flow needs to pass all $\Psi$ VNFs to complete the service. Here, any $N\geq1$ servers allocate one sub-SFCs to serve one sub-flow, whereby any server $s\in[1,N]$ of a certain sub-flow contains $\psi_s\leq\Psi$, $\sum_{s=1}^{N}\psi_s=\Psi$, different types of VNFs. To ensure service success, $r$ redundant network components can be utilized in case some of $k$ main components fail, where $r\leq k$. We denote the reliability values of main and redundant servers and VM/VNFs as $\varphi$, $\varphi_r$ and $\upsilon$, $\upsilon_r$, respectively, and assume that these values are the same for all servers and VNFs. Moreover, we assume that there exists a connection, i.e., a path segment, to certain main or redundant server out of all alternative connections to the same server with probability for connectivity $\phi$ or $\phi_r$, respectively. However, the value $\phi$ and $\phi_r$ are the same for connectivity to any main and redundant servers, respectively. Since any sub-SFC is distributed over $N$ servers, we need at least $k(N+1)$ main- and $rN$ redundant path segments. For the analysis below, we assume that the failures of different components occur independently.  

Without any failure protection, all path segments to any server, all servers and VNFs have to be available. Here, $kN$ servers allocate $k$ sub-SFCs to serve $k$ sub-flows requiring $k(N\!\!+\!\!1)$ available paths segments. Thus, the service success is
\begin{equation}\label{PureSuccess}
R_{0}(k)=\big[\phi^{N+1}\varphi^N\prod_{s=1}^{N}\upsilon^{\psi_s}\big]^k
\end{equation}

\subsection{Backup protection}
To provide service reliability, i.e., protection of $k$ active sub-SFCs, there are $r$ backup (redundant) sub-SFCs allocated on $r N$ backup servers. In case of failures, the backup VNFs can replace failed active VNFs of the same type. 
All active and backup servers in DCN need to be connected by at least one available path segment, whereby the path segment from the last server to destination end-host within DCN has to be always available. In contrast, a failure of any other path segment to active or backup servers results in traffic redirection to a reachable backup server. In this case, the path segment to a certain backup server needs to be available. In other words, when a server can not be reached due to connectivity failure, the reachable backup server replaces this unreachable server.


For example, assume that only VNFs can fail with probability $(1-\upsilon)$. The service success is then a probability that at least $k$ over all $k+r$ VNFs of any type do not fail and can serve $k$ sub-flows. Thus, at most $r$ VNFs of any type can fail and service success is
$R=[\sum_{i=0}^{r}\binom{r+k}{i}(1-\upsilon)^i\upsilon^{r+k-i}]^{\Psi}$, where and $\sum_{i=x}^{X} a_i=0\text{, if } x>X$.
Using this general expression and approach, we next derive the service success provided by backup protection.

Since the allocation of backup sub-SFCs follows the allocation principles of active sub-SFCs by design, $r$ backup sub-SFCs are distributed over $(r\cdot N)$ different servers, i.e., over $N$ backup servers each. The $k$ path segments to the destination within DCN have to be available with probability $\phi^k$. Let us consider all $k$ active and $r$ backup servers of type $s$, which contain $\psi_s$ VNFs. Generally, at most $\xi=r$ connections to active servers can fail resulting in unavailability of $r$ active servers and of all $r\psi_s$  VNFs on these servers, though without an impact on service success, if all redundant backup components are available. If $\xi$ path segments failed, the service can be still completed, when at least $\xi$ path segments to $\xi$ backup servers and $\xi\cdot\psi_s$ backup VNFs of certain type are available. Thus, at most $\gamma=r-\xi$ path segments to backup servers and $f=r-\xi-\gamma$ other active servers and at most $l=r-\xi-\gamma-f$ backup servers can additionally fail without an impact on the resulting service success. 
On all active and backup servers, at most $i=r-\xi-\gamma-f-l$ active and, then, at most $j=r-\xi-\gamma-f-l-i$ backup VNFs are allowed to fail. As a result, the service success is derived as follows
\begin{small}
\begin{equation}\label{multibackup}
\begin{split}
&R_b\!\!=\phi^k\!\!\prod_{s=1}^{N}\!\sum_{\xi=0}^{r}\binom{k}{\xi}\phi^{k-\xi}(1-\phi)^{\xi}\!\sum_{\gamma=0}^{r-\xi}\binom{r}{\gamma}\phi_r^{r-\gamma}(1-\phi_r)^{\gamma}
\\&\sum_{f=0}^{r-\xi-\gamma}\!\!\!\binom{k-\xi}{f}\varphi^{k-f-\xi}(1-\varphi)^f\!\!\!\!\!\!\sum_{l=0}^{r-\xi-\gamma-f}\!\!\!\binom{r\!-\!\gamma}{l}\varphi_r^{r\!-\!\gamma-l}\\&(1-\varphi_r)^l
\bigg[\sum_{i=0}^{r-\xi-\gamma-f-l}\!\!\!\binom{k-\xi-f}{i}(1-\upsilon)^i
 \upsilon^{k-\xi-f-i} 
 \\&\sum_{j=0}^{r-\xi-\gamma-f-l-i}
\binom{r-\gamma-l}{j}(1-\upsilon_r)^j \upsilon_r^{r-\gamma-l-j}\bigg]^{\psi_s}
\end{split}
\end{equation}
\end{small}
, where the first sum describes possible connectivity loss to active servers, the second sum describes the connectivity loss to backup servers, the third and forth sums consider possible failures of remaining available active and backup servers, respectively; and the fifth and sixth sums consider the failure of active and backup VNFs available on connected active and backup servers, respectively. 
\begin{table}[]
\centering
\caption{Notation}\label{t1}
\vspace{-0.3cm}
\resizebox{\columnwidth}{!}{
\begin{tabular}{|l|l|}
\cline{1-2}
  $\varphi$, $\varphi_r$& reliability of the active, redundant server, respectively; \\ 
 $\upsilon$, $\upsilon_r$&  reliability of VNF on active, redundant server, respectively;  \\
 $\phi$, $\phi_r$&  probability for connectivity to main, redundant server, respectively;  \\
 $\Psi$ &   number of VNFs in a SFC and, thus, sub-SFC; \\
$k$ &   number of parallel sub-flows and, thus, active sub-SFCs; \\
 $N$&  number of active servers per sub-flow;  \\
 $\psi_s$&  number of VNFs allocated on server $s\in[1,N]$;  \\
 $r$& number of redundant/backup sub-SFCs;\\
  \cline{1-2}
\end{tabular}
}\vspace{-0.5cm}
\end{table}
\subsection{Protection with erasure coding}
In case of header processing only, i.e., SFC consists of h-VNFs only, protection with systematic erasure coding can be applied. Then, the service reliability $R_h$ or $R^{'}_{h}$ of any main or redundant sub-flow is generally independent from other parallel coded sub-flows and is only the function of connectivity, server and VNF reliability, i.e., $\phi$, $\varphi$ and $\upsilon$ or $\phi_r$, $\varphi_r$ and $\upsilon_r$, respectively. 
\begin{equation}\label{RH}
R_h=\varphi^N\upsilon^\Psi \phi^{N+1} \text{  or  }R^{'}_{h}=\varphi_r^N\upsilon_r^\Psi \phi_r^{N+1}
\end{equation}

Without coding and VNF redundancy, the service success can be determined by Eq. \eqref{PureSuccess}.  When $r$ redundant sub-flows are generated by encoding and sent over $r$ parallel redundant paths through $r$ parallel redundant sub-SFCs, the service success can be calculated with Eq. \eqref{DecSuccess} as discussed below. In contrast to backup protection, the connectivity to destination within DCN (decoder) can fail with probability $(1-\phi)$ or $(1-\phi_r)$ without degrading service success. The failures of connectivity will result in acceptable traffic loss, whereby any lost sub-flow can be recovered by decoding. The service success is a probability that at least $k$ sub-flows out of $(k+r)$ parallel coded sub-flows can reach the decoder and calculated as follows
\begin{equation}\label{DecSuccess}
R_c\!=\!\!\!\sum_{i=0}^{r}\binom{k}{i}R_h^{k-i}(1\!-\!R_h)^i\!\sum_{j=0}^{r-i}\binom{r}{j}{R^{'}_{h}}^{r-j}(1\!-\!R^{'}_{h})^j
\end{equation}
, where the first sum consider possible loss of main sub-flows and the second sum describes the possible affordable loss of the remaining redundant sub-flows. Both losses, however, are the result of server or VNF failure or connectivity loss regarding Eq. \eqref{RH}.

In case of payload processing, the decoding is only possible, if $k$ out of $k+r$ sub-flows are not interrupted and reach the decoder. Thus, the service success probability depends on both $R_c$ and $R_b$. Let us assume that any SFC can be split into multiple parts, i.e., $M_\text{h}$ chained parts, where $\Psi_{\text{h},i}$, $1\leq i\leq M_\text{h}$, VNFs can work with coded packets and process GCH, and $M_\text{p}$ chained parts with $\Psi_{\text{p},j}$, $1\leq j\leq M_\text{p}$ VNFs, which require prior decoding and backup protection. Then, the total number of different VNFs in any SFC is $\Psi=\sum_{i=1}^{M_\text{h}}\Psi_{\text{h},i} + \sum_{j=1}^{M_\text{p}}\Psi_{\text{p},j}$. The service success in case of hybrid protection can be derived as
$R=R_b^{M_\text{p}}\cdot R_c^{M_\text{h}}$.

\subsection{Analysis of Overhead}
For the method of backup protection we define the overhead as probability of VNF migration and traffic redirection, while for coding protection the overhead is defined as  the probability that the decoding is needed.
In backup protection, each failed VNF of certain type can be replaced by any out of $r$ backup VNFs of the same type through redirection of individual sub-flows. Thus, the traffic redirection is required, if at least one active VNF is not available. Let us derive the probability that all $k$ active sub-SFCs are available as $\prod_{s=1}^{N}(\phi\varphi\upsilon^\psi_s)^k$, then the probability for traffic redirection is calculated as
\begin{equation}\label{Redirect}
P_\text{R}=1-\prod_{s=1}^{N}(\phi\varphi\upsilon^\psi_s)^k
\end{equation}

In case of coding protection, and since we apply systematic erasure code, where $k$ sub-flows are unchanged after encoding and $r$ sub-flows present a linear combination of the $k$ original sub-flows, the decoding is only required if at least one out of $k$ unchanged sub-flows is lost. There is no need for decoding, when $f\le r$ redundant sub-flows were lost.
Thus, the probability for decoding need can be derived as
\begin{equation}\label{Dec2}
P_\text{dec}\!\!=\!\!\!\sum_{f=1}^{r}\!\sum_{i=0}^{r-f}\binom{k}{f}R_h^{k-f}(1-R_h)^f \binom{r}{i}{R^{'}_{h}}^{r-i}(1-R^{'}_{h})^i  
\end{equation}
To decrease the probability for decoding need $P_\text{dec}$, the $k$ unchanged sub-flows can be sent over more reliable routes and $r$ redundant modified sub-flows over less reliable routes.

\section{Analytical Results}
We now evaluate the service success and overhead of the proposed VNF protection methods. We verify analytical results by Monte-Carlo simulations, and find that the simulation results overlapped with analytical results (with 95\% confidence interval) and are not shown for clarity. In all results, we assume that VNF failure is more likely than connectivity or server failures, and set $N = 2$ and $\Psi=4$.
 \begin{figure}[!t]
\centering
\includegraphics[width=1\columnwidth]{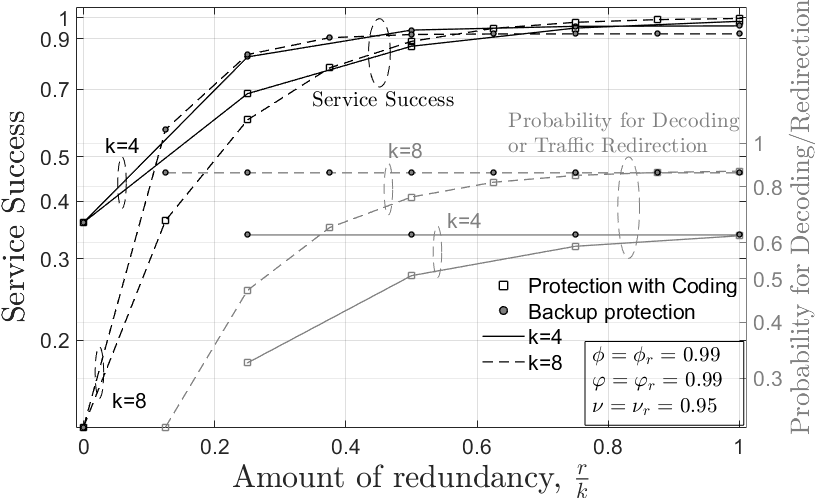}%
  \vspace{-0.2cm}
  \caption{Service success and Probability for Overhead vs number of parallel sub-flows and redundant sub-SFCs.}
\vspace{-0.4cm}
  \label{KR}
\end{figure}
\par Fig. \ref{KR} shows the service success and the overhead probability, as a function of the number of main and redundant parallel sub-SFCs, i.e., $k$ and $r$. In contrast to backup protection, which provides high service success with a few redundant sub-SFCs, the protection with coding outperforms backup protection, if there are $\geq 60\%$ redundant sub-SFCs. Service success can be increased by increasing $k$ and $r$. The overhead probability of backup protection is independent from the amount of backups and increases with increasing $k$. The probability of decoding is less than the probability for traffic redirection, however, increases with increasing $k$ and $r$.
\begin{figure}[!t]
\centering
\includegraphics[width=0.95\columnwidth]{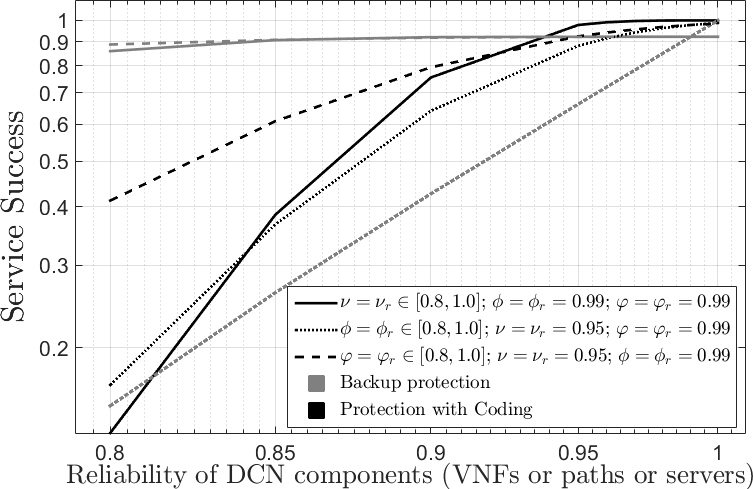}%
  \vspace{-0.2cm}
  \caption{Service success in case of $k=8$ parallel sub-flows and $r=6$ redundant sub-SFCs vs. Reliability of DCN components}
\vspace{-0.4cm}
  \label{RComponent}
\end{figure}
\par Fig. \ref{RComponent} shows the service success of both VNF protection methods as a function of reliability of certain network components, i.e., VNFs ($\nu=\nu_r$) or path segments ($\phi=\phi_r$) or servers ($\varphi=\varphi_r$). The service success of backup protection is almost independent from reliability value of VNFs and servers, and increases with increasing connectivity. The service success of coding protection increases with increasing component reliability and outperforms backup protection for any values of connectivity ($\phi=\phi_r$) and when the reliability of VNFs or servers is higher than $\approx 0.95$.
 \begin{figure}[!t]
\centering
\includegraphics[width=1\columnwidth]{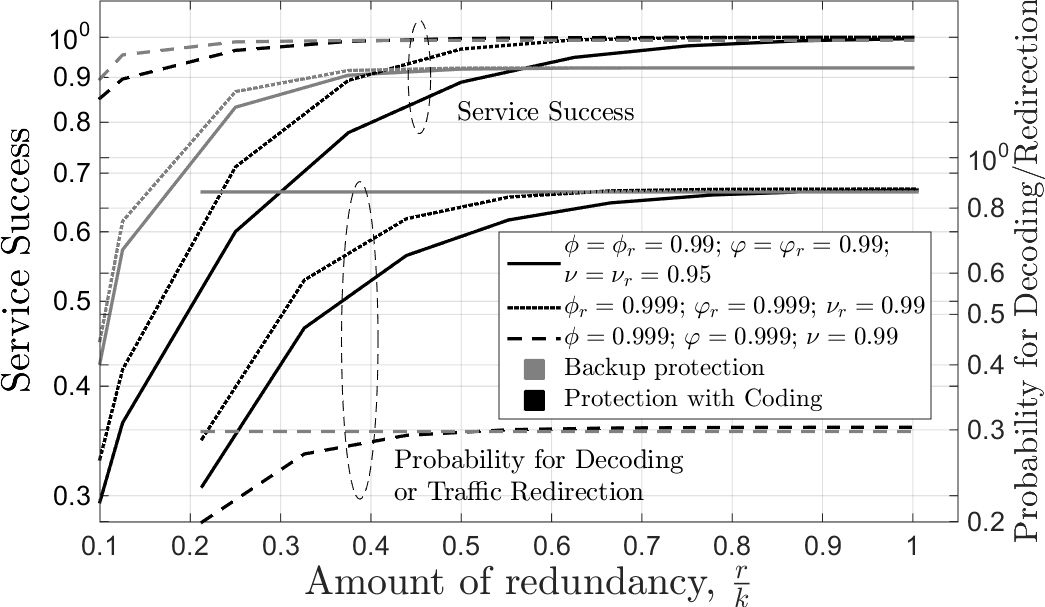}%
  \vspace{-0.2cm}
  \caption{Service success and Probability for Overhead for $k=8$ parallel sub-flows  vs number of redundant sub-SFCs.  Solid lines: Main and redundant components are equally reliable; doted lines: Redundant components are more reliable; dashed lines: Main components are more reliable.}
\vspace{-0.4cm}
  \label{RRcompare}
\end{figure}
\par Fig. \ref{RRcompare} shows the service success and the overhead probability as a function of the number of redundant parallel sub-SFCs $r$. Here, we compare three scenarios, where the main components are equally, less or more reliable than the redundant components. In all scenarios, the coding protection outperforms the backup protection regarding the service success in case of large redundancy, i.e., $r> 0.5$. In general, the third studied scenario, i.e., $\nu>>\nu_r$, $\phi>>\phi_r$, $\varphi>>\varphi_r$, provides the highest service success and the lowest overhead probability. The lowest service success and the highest overhead is reached in the first and second scenario, respectively.
 \begin{figure}[!t]
\centering
\includegraphics[width=1\columnwidth]{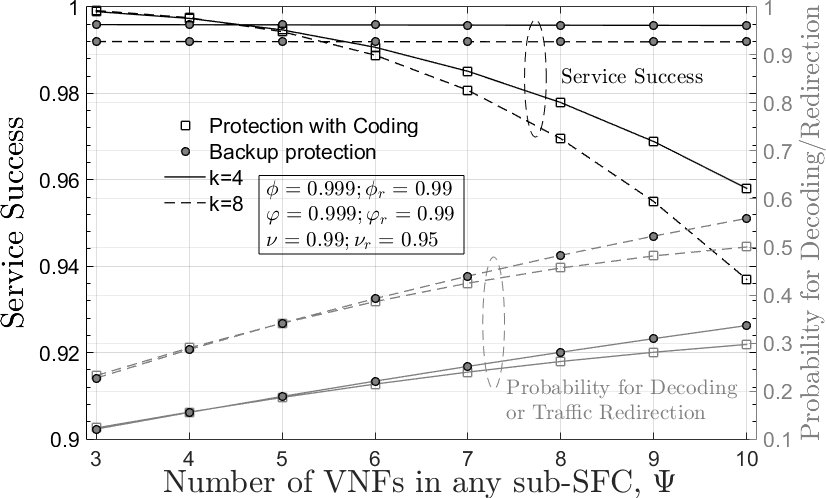}%
  \vspace{-0.2cm}
  \caption{Service success and Probability for Overhead for $k=4$ and $k=8$ parallel sub-flows  and $r=3$ and $r=4$ redundant sub-SFCs, respectively, vs number of VNFs in any sub-SFC.}
 \vspace{-0.5cm}
  \label{VNFnum}
\end{figure}
\par Fig. \ref{VNFnum} illustrates the service success and probability for overhead as a function of the number of VNFs in any sub-SFC ($\Psi$) and the number of main parallel sub-SFCs. Here, we assumed a scenario, where the main components are more reliable than the redundant components, i.e., $\nu>>\nu_r$, $\phi>>\phi_r$, $\varphi>>\varphi_r$. The service success in case of backup protection is independent from $\Psi$ and increases with decreasing $k$. In contrast, the service success decreases with increasing $\Psi$ in case of coding protection. The overhead probability increases with $\Psi$ for both protection methods and decreases with decreasing $k$, while coding protection outperforms backup protection.
 \begin{figure}[!t]
\centering
\includegraphics[width=1\columnwidth]{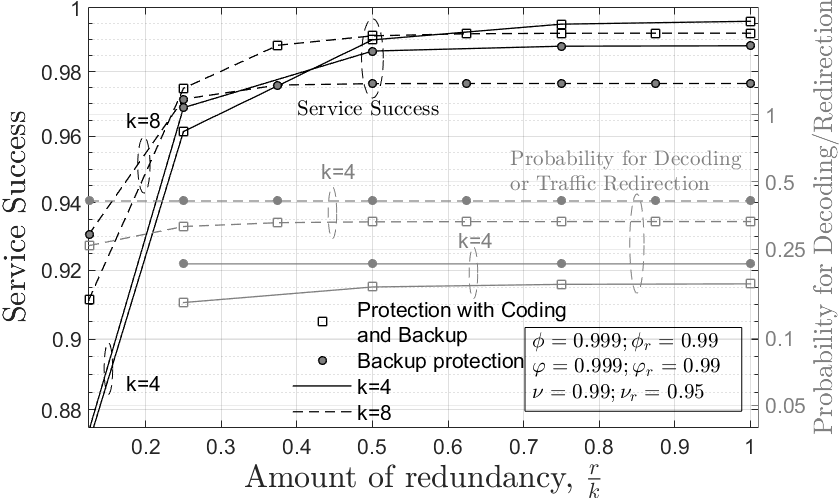}%
  \vspace{-0.2cm}
  \caption{Service success and Probability for Overhead vs number of parallel sub-flows and redundant sub-SFCs.}
\vspace{-0.4cm}
  \label{hybridRes}
\end{figure}
\par Finally, Fig. \ref{hybridRes} shows the service success and the overhead probability  for the use case from Fig. \ref{hybridPlace}, whereby we change the amount of redundancy (backups) and compare the backup protection of sub-SFCs with hybrid protection, i.e., coding protection of VNFs1-VNFs3 and VNF5 and backup protection of VNFs4. The hybrid protection outperforms the pure backup protection for any $k$ providing higher service success and lower probability for decoding need, when the amount of redundant resources is larger than 25\%. When amount of redundancy is less than $\approx 0.4$ the highest service success can be reach with high level of flow parallelization $k=8$. In contrast, the low level of flow parallelization $k=4$ results in high service reliability, when the amount of redundancy is $\geq 0.4$. The minimal overhead can be reached, however, for $k=4$ for both protection methods. 

\section{Conclusion}
We analytically studied and compared backup failure protection and failure protection with systematic erasure coding regarding the resulting service success and required overhead. Based on the probability theory and by assuming that the reliability of the individual VMs, servers and path segments are known, we derived analytical expressions for evaluation of the service success and the probability for additional overhead such as VNF migration or the need for decoding. The results showed that the highest end-to-end service success can be reached, when the main active components (paths segments, servers, VNFs) have the higher reliability than the utilized redundant or backup components. The backup protection shows benefits, when the number of VNFs in any sub-SFC is high, e.g., more than 5 in studied use case. In contrast, when any sub-SFC contains few VNFs, the coding protection provides higher service success while requiring less reliable path segments than backup protection. Moreover, coding protection showed lower probability for decoding, as compared to probability for VNF migration and traffic redirection in case of backup protection. We also showed, that a combination of both methods increases the service success  and decreases the overhead significantly.

%


\ifCLASSOPTIONcaptionsoff
  \newpage
\fi


\end{document}